\documentclass[aps,prl,superscriptaddress,showpacs,floatfix,twocolumn]{revtex4}

\usepackage{graphicx}   
\def \gevc {GeV/$c$}                                 

\begin{document}

\title{Modifications to Di-Jet Hadron Pair Correlations in Au+Au 
Collisions at $\sqrt{s_{NN}}=$200~GeV
}

\newcommand{\abilene}{Abilene Christian University, Abilene, TX 79699, USA}
\newcommand{\acadsin}{Institute of Physics, Academia Sinica, Taipei 11529, Taiwan}
\newcommand{\banaras}{Department of Physics, Banaras Hindu University, Varanasi 221005, India}
\newcommand{\barc}{Bhabha Atomic Research Centre, Bombay 400 085, India}
\newcommand{\bnl}{Brookhaven National Laboratory, Upton, NY 11973-5000, USA}
\newcommand{\caucr}{University of California - Riverside, Riverside, CA 92521, USA}
\newcommand{\ciae}{China Institute of Atomic Energy (CIAE), Beijing, People's Republic of China}
\newcommand{\cns}{Center for Nuclear Study, Graduate School of Science, University of Tokyo, 7-3-1 Hongo, Bunkyo, Tokyo 113-0033, Japan}
\newcommand{\columbia}{Columbia University, New York, NY 10027 and Nevis Laboratories, Irvington, NY 10533, USA}
\newcommand{\dapnia}{Dapnia, CEA Saclay, F-91191, Gif-sur-Yvette, France}
\newcommand{\debrecen}{Debrecen University, H-4010 Debrecen, Egyetem t{\'e}r 1, Hungary}
\newcommand{\fsu}{Florida State University, Tallahassee, FL 32306, USA}
\newcommand{\gsu}{Georgia State University, Atlanta, GA 30303, USA}
\newcommand{\hiroshima}{Hiroshima University, Kagamiyama, Higashi-Hiroshima 739-8526, Japan}
\newcommand{\ihepprot}{IHEP Protvino, State Research Center of Russian Federation, Institute for High Energy Physics, Protvino, 142281, Russia}
\newcommand{\isu}{Iowa State University, Ames, IA 50011, USA}
\newcommand{\jinrdubna}{Joint Institute for Nuclear Research, 141980 Dubna, Moscow Region, Russia}
\newcommand{\kaeri}{KAERI, Cyclotron Application Laboratory, Seoul, South Korea}
\newcommand{\kangnung}{Kangnung National University, Kangnung 210-702, South Korea}
\newcommand{\kek}{KEK, High Energy Accelerator Research Organization, Tsukuba, Ibaraki 305-0801, Japan}
\newcommand{\kfki}{KFKI Research Institute for Particle and Nuclear Physics of the Hungarian Academy of Sciences (MTA KFKI RMKI), H-1525 Budapest 114, POBox 49, Budapest, Hungary}
\newcommand{\korea}{Korea University, Seoul, 136-701, Korea}
\newcommand{\kurchatov}{Russian Research Center ``Kurchatov Institute", Moscow, Russia}
\newcommand{\kyoto}{Kyoto University, Kyoto 606-8502, Japan}
\newcommand{\labllr}{Laboratoire Leprince-Ringuet, Ecole Polytechnique, CNRS-IN2P3, Route de Saclay, F-91128, Palaiseau, France}
\newcommand{\lawllnl}{Lawrence Livermore National Laboratory, Livermore, CA 94550, USA}
\newcommand{\losalamos}{Los Alamos National Laboratory, Los Alamos, NM 87545, USA}
\newcommand{\lpc}{LPC, Universit{\'e} Blaise Pascal, CNRS-IN2P3, Clermont-Fd, 63177 Aubiere Cedex, France}
\newcommand{\lund}{Department of Physics, Lund University, Box 118, SE-221 00 Lund, Sweden}
\newcommand{\muenster}{Institut f\"ur Kernphysik, University of Muenster, D-48149 Muenster, Germany}
\newcommand{\myongji}{Myongji University, Yongin, Kyonggido 449-728, Korea}
\newcommand{\nagasaki}{Nagasaki Institute of Applied Science, Nagasaki-shi, Nagasaki 851-0193, Japan}
\newcommand{\newmex}{University of New Mexico, Albuquerque, NM 87131, USA}
\newcommand{\nmsu}{New Mexico State University, Las Cruces, NM 88003, USA}
\newcommand{\ornl}{Oak Ridge National Laboratory, Oak Ridge, TN 37831, USA}
\newcommand{\orsay}{IPN-Orsay, Universite Paris Sud, CNRS-IN2P3, BP1, F-91406, Orsay, France}
\newcommand{\pnpi}{PNPI, Petersburg Nuclear Physics Institute, Gatchina, Russia}
\newcommand{\riken}{RIKEN, The Institute of Physical and Chemical Research, Wako, Saitama 351-0198, Japan}
\newcommand{\rikjrbrc}{RIKEN BNL Research Center, Brookhaven National Laboratory, Upton, NY 11973-5000, USA}
\newcommand{\saispbstu}{Saint Petersburg State Polytechnic University, St. Petersburg, Russia}
\newcommand{\saopaulo}{Universidade de S{\~a}o Paulo, Instituto de F\'{\i}sica, Caixa Postal 66318, S{\~a}o Paulo CEP05315-970, Brazil}
\newcommand{\seoulnat}{System Electronics Laboratory, Seoul National University, Seoul, South Korea}
\newcommand{\stonybrkc}{Chemistry Department, Stony Brook University, SUNY, Stony Brook, NY 11794-3400, USA}
\newcommand{\stonycrkp}{Department of Physics and Astronomy, Stony Brook University, SUNY, Stony Brook, NY 11794, USA}
\newcommand{\subatech}{SUBATECH (Ecole des Mines de Nantes, CNRS-IN2P3, Universit{\'e} de Nantes) BP 20722 - 44307, Nantes, France}
\newcommand{\tenn}{University of Tennessee, Knoxville, TN 37996, USA}
\newcommand{\titech}{Department of Physics, Tokyo Institute of Technology, Tokyo, 152-8551, Japan}
\newcommand{\tsukuba}{Institute of Physics, University of Tsukuba, Tsukuba, Ibaraki 305, Japan}
\newcommand{\vandy}{Vanderbilt University, Nashville, TN 37235, USA}
\newcommand{\waseda}{Waseda University, Advanced Research Institute for Science and Engineering, 17 Kikui-cho, Shinjuku-ku, Tokyo 162-0044, Japan}
\newcommand{\weizmann}{Weizmann Institute, Rehovot 76100, Israel}
\newcommand{\yonsei}{Yonsei University, IPAP, Seoul 120-749, Korea}
\affiliation{\abilene}
\affiliation{\acadsin}
\affiliation{\banaras}
\affiliation{\barc}
\affiliation{\bnl}
\affiliation{\caucr}
\affiliation{\ciae}
\affiliation{\cns}
\affiliation{\columbia}
\affiliation{\dapnia}
\affiliation{\debrecen}
\affiliation{\fsu}
\affiliation{\gsu}
\affiliation{\hiroshima}
\affiliation{\ihepprot}
\affiliation{\isu}
\affiliation{\jinrdubna}
\affiliation{\kaeri}
\affiliation{\kangnung}
\affiliation{\kek}
\affiliation{\kfki}
\affiliation{\korea}
\affiliation{\kurchatov}
\affiliation{\kyoto}
\affiliation{\labllr}
\affiliation{\lawllnl}
\affiliation{\losalamos}
\affiliation{\lpc}
\affiliation{\lund}
\affiliation{\muenster}
\affiliation{\myongji}
\affiliation{\nagasaki}
\affiliation{\newmex}
\affiliation{\nmsu}
\affiliation{\ornl}
\affiliation{\orsay}
\affiliation{\pnpi}
\affiliation{\riken}
\affiliation{\rikjrbrc}
\affiliation{\saispbstu}
\affiliation{\saopaulo}
\affiliation{\seoulnat}
\affiliation{\stonybrkc}
\affiliation{\stonycrkp}
\affiliation{\subatech}
\affiliation{\tenn}
\affiliation{\titech}
\affiliation{\tsukuba}
\affiliation{\vandy}
\affiliation{\waseda}
\affiliation{\weizmann}
\affiliation{\yonsei}
\author{S.S.~Adler}	\affiliation{\bnl}
\author{S.~Afanasiev}	\affiliation{\jinrdubna}
\author{C.~Aidala}	\affiliation{\bnl}
\author{N.N.~Ajitanand}	\affiliation{\stonybrkc}
\author{Y.~Akiba}	\affiliation{\kek} \affiliation{\riken}
\author{J.~Alexander}	\affiliation{\stonybrkc}
\author{R.~Amirikas}	\affiliation{\fsu}
\author{L.~Aphecetche}	\affiliation{\subatech}
\author{S.H.~Aronson}	\affiliation{\bnl}
\author{R.~Averbeck}	\affiliation{\stonycrkp}
% ...
\collaboration{PHENIX Collaboration} \noaffiliation

\date{\today}


\begin{abstract}
Azimuthal correlations of high-$p_T$ charged hadron
pairs from (di-)jet-fragmentation are studied at mid-rapidity in Au+Au 
collisions at $\sqrt{s_{_{\rm NN}}}$=200~GeV. 
The distribution of jet-associated partner hadrons 
(1.0$<p_{T}<$2.5~\gevc) per trigger hadron (2.5$<p_{T}<$4.0~\gevc) 
is found to vary 
with collision centrality, both in shape and in yield,
indicating a significant effect of the nuclear collision medium on 
the \mbox{(di-)jet} fragmentation process.  

\end{abstract}



\maketitle



Energetic collisions between heavy ions at the 
Relativistic Heavy Ion Collider (RHIC) have 
been shown to produce matter with extremely 
high energy density~\cite{whitepaper}.
This matter has been observed to strongly suppress the yield
of hadrons with large transverse momenta in central Au+Au
collisions, compared to yields in p+p collisions scaled by
the number of binary nucleon-nucleon collisions~\cite{ppg003,
ppg014,star_supp,ppg023}.  Such a suppression was predicted to result
from energy loss of hard-scattered partons (light quarks and gluons) 
traversing the dense matter prior to forming the observed 
hadrons~\cite{wang,bjorken}. 
If the parton encounters a sufficient amount of
dense matter, the energy loss could strongly modify 
its fragmentation into jets of hadrons.

Strong suppression of the away-side jet has been observed at 
RHIC~\cite{starb2b}.  However, it is unclear at present how the
lost energy is transported by the dense medium, and how the 
parton-medium interaction affects the fragmentation process. 
Recently, there have been predictions that the coupling 
of jets to a strongly interacting medium may modify the angular 
distribution and number of jet 
fragments~\cite{Fries,Wiedemann,Ko,Hwa,Majumder:2004pt,
cherenkov_1,cherenkov_2,cherenkov_3}.  Quarks from hard 
scattering processes may recombine with thermal quarks from 
the dense medium\cite{Ko,Hwa}.  Co-moving radiated gluons may 
produce a ``wake'' in the medium, further increasing the number of
quarks available for building hadrons in the jet fragmentation 
process\cite{Fries,Wiedemann}.  It has even been proposed that
the energy deposited in the medium creates a shock wave around 
the propagating parton, thereby creating a ``conical flow''
akin to a sonic boom in a 
fluid~\cite{cherenkov_1,cherenkov_2,cherenkov_3}. 
To investigate the transport
of lost parton energy, the PHENIX experiment at RHIC measures 
azimuthally correlated hadrons arising from jet fragmentation 
as a function of 
centrality in Au+Au collisions.   Such studies, in effect, use
hard scattered partons as short-wavelength probes of the
produced medium.


The analysis presented in this Letter uses data from 
$\sqrt{s_{_{NN}}}$=200~GeV Au+Au collisions in the 
PHENIX \mbox{Run-2} data set.   
Charged particles are reconstructed in the central arms of
PHENIX using drift chambers, each with azimuthal coverage
of $\pi/2$, and two layers of multi-wire proportional chambers 
with pad readout (PC1, PC3)~\cite{nim_phenix}.  Pattern recognition
is based on a combinatorial Hough transform in the track bend
plane, with the polar angle determined by PC1 and 
the collision vertex along the beam direction\cite{Mitchell_nim}.
Particle momenta are measured with a resolution
$\delta p/p = 0.7\% \oplus 1.0\%p~($\gevc$)$.
To reject most background from albedo, conversions,
and decays, a confirmation hit is required within
a 2$\sigma$ matching window in PC3~\cite{ppg003}. 
The Au+Au event centrality is determined
using the PHENIX beam-beam counters (BBC) and 
zero-degree calorimeters (ZDC)~\cite{ppg009}.


The traditional identification of jets through hadronic 
calorimetry and cluster algorithms
is problematic in A+A collisions at RHIC, since 
low-energy jets ($<$10-20~GeV) are overwhelmed by 
other produced particles in the underlying event 
and high-energy jets are 
relatively rare at $\sqrt{s_{_{\rm NN}}}$=200~GeV.  
Instead, we study hard-scattered single partons 
and parton pairs through angular correlations of 
high-$p_{T}$ hadron pairs.  We examine the distribution 
of pairs over relative azimuthal angle, $dN^{AB}/d(\Delta\phi)$, 
where $A$ and $B$ denote charged particles 
in the PHENIX pseudorapidity 
acceptance ($|\eta|<0.35$) and in $p_{T}$ bins
2.5~\gevc $<p_{T}^{A}<$4.0~\gevc $ $ (``trigger''),
and 1.0~\gevc $<p_{T}^{B}<$2.5~\gevc $ $ (``partner'').
Pairs from fragments of the same jet are expected 
to appear near $\Delta\phi \sim 0$, 
while $\Delta\phi\sim\pi$ indicates one hadron each from the 
outgoing hard-scattered parton pair (hereafter \mbox{``dijet''}).


The PHENIX acceptance at central rapidity is non-uniform 
in azimuth. We correct for the shape of the acceptance in 
$\Delta\phi$ by constructing an area normalized correlation 
function, utilizing pairs from mixed events:
\begin{equation}
C(\Delta \phi) \equiv 
\frac{Y^{AB}_{\rm Same}(\Delta\phi)}
                            {Y^{AB}_{\rm Mixed}(\Delta\phi)}
                            \times
\frac{\int Y^{AB}_{\rm Mixed}(\Delta\phi)}
                            {\int Y^{AB}_{\rm Same}(\Delta\phi)}
\propto  \frac{dN^{AB}}{d(\Delta\phi)}                         
\label{Eq:CF_defined}
\end{equation}
\noindent
where $Y^{AB}_{\rm Same}(\Delta\phi)$ and 
$Y^{AB}_{\rm Mixed}(\Delta\phi)$ are, respectively,
the uncorrected yields of pairs in the same and in mixed
events within each data sample.


\begin{figure}[thb]
\includegraphics[width=1.0\linewidth]{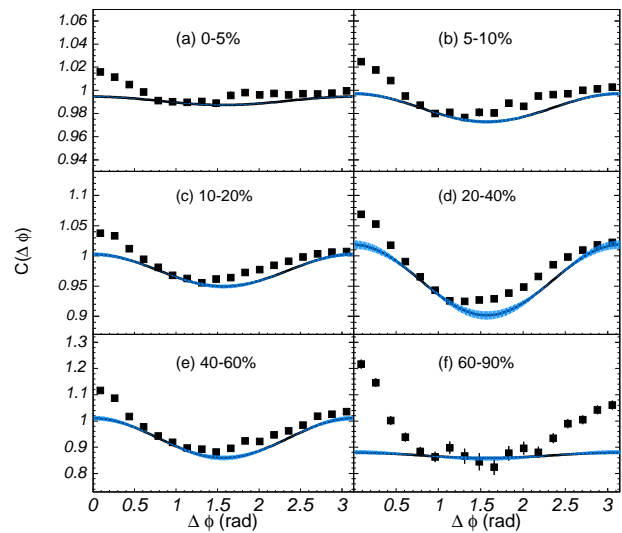}  
\caption{Correlation functions, $C(\Delta\phi)$, 
for pairs of charged hadrons with 2.5~\gevc $<p_{T}^{A}<$4.0~\gevc $ $
and 1.0~\gevc $<p_{T}^{B}<$2.5~\gevc $ $ in different bins
of collision centrality.  The solid bands indicate the estimate 
of the background pair component (see text) within one unit 
of its systematic error.}
\label{fig:Unsubtracted}
\end{figure}


The correlation functions are shown in Fig.~\ref{fig:Unsubtracted}, 
folded into the range $0<\Delta\phi<\pi$.
For the most peripheral collisions (cf. Fig.~\ref{fig:Unsubtracted}f),
the correlation function shows two well-defined peaks 
centered at $\Delta\phi=0$ and $\Delta\phi=\pi$, which we can
attribute to (di-)jet pairs.  For more central collisions, 
there is a similar peak at $\Delta\phi=0$, a broader peak 
at $\Delta\phi=\pi$, and the apparent minima appear at
$\Delta\phi < \pi/2$. 
These features reflect a mixture of (di-)jet pairs 
and underlying events with particle flow along
the reaction plane~\cite{methods}.


In order to extract and examine the jet-induced pairs
we analyze the pair distribution in the context of a 
{\em two-source model}, assuming that each hadron can be attributed 
to either (i)~a jet fragmentation source, 
or (ii)~the underlying event.
Additionally we assume that the $\phi$ distributions 
for $A$ or $B$ inclusive single particles, summed over both 
sources, have a shape proportional to 
$(1+2 \langle v_{2}^{A \, {\rm or} \, B}\rangle \cos(2(\phi-\Phi_{\rm RP})))$
relative to azimuthal angle $\Phi_{\rm RP}$ of the reaction 
plane of each event.  All pairs which are not from the same 
jet or dijet fragmentation are termed {\em background pairs},
and are taken to have no angular correlation beyond 
having their distributions respect the same reaction plane.  
In principle, contributions from resonance decays and global
transverse momentum conservation can also affect the distribution 
of background pairs, but we estimate these effects to be 
negligible for these $p_{T}$ ranges and the PHENIX $\eta$ acceptance.   
The distribution of background pairs over
$\Delta\phi$ is then proportional to 
$(1+2 \langle v_{2}^{A} v_{2}^{B} \rangle \cos(2\Delta\phi))$.


Given a normalization, $C(\Delta\phi)$ can then 
be decomposed into two pieces, one proportional to the distribution 
of background pairs and another $J(\Delta \phi)$
proportional to that of the (di-)jet pairs:

\begin{equation}
C(\Delta\phi) = 
b_{0} (1+2 \langle v_{2}^{A} v_{2}^{B} \rangle \cos(2\Delta\phi))
+ J(\Delta\phi)
\label{Eq:CF_decomp}
\end{equation}
\noindent
We approximate $\langle v_{2}^{A} v_{2}^{B} \rangle =
\langle v_{2}^{A} \rangle \langle v_{2}^{B} \rangle$, 
and we measure $\langle v_{2}^{A} \rangle$ and 
$\langle v_{2}^{B} \rangle$ for each centrality and particle $p_T$ 
bin through a standard reaction-plane analysis 
using the PHENIX BBC to reconstruct 
the reaction plane event-by-event. The large rapidity gap, 
$\Delta \eta > 2.75$, between the central arm acceptance 
and the BBC acceptance substantially reduces non-flow contributions 
to the measured $v_2$ values, particularly those arising from di-jets.
The results are shown in Table~\ref{table:v2_tab}; they are 
consistent with prior PHENIX $v_2$ measurements~\cite{ppg022}.
%
%
\begin{table}[hbpt]
\caption{\label{table:v2_tab}
Anisotropy values for bins 
$A$~($2.5<p_T <4.0$ GeV/c) and $B$~($1.0<p_T <2.5$ GeV/c),
shown with statistical errors, and values of 
$\phi_{Min}$ (see text).  The relative systematic 
errors on the anisotropies are estimated to be $\pm$6\%
for the five most central samples and $\pm$40\% for the
most peripheral sample.  The systematic
errors on the $v_{2}$'s are dominated by the uncertainty
in the correction for reaction plane 
resolution\cite{ppg022}, and we assume them
to be completely correlated between the two $p_{T}$
bins in each centrality sample.}
\begin{ruledtabular} \begin{tabular}{lllc}
Centrality ({\%})& $\left\langle v_2^B \right\rangle$ 
	& $\left\langle v_2^A \right\rangle$
	& $\phi_{Min}$ (rad)    \\
\hline
0-5 & 0.035 $\pm$ 0.001 &  0.052 $\pm$ 0.007 & 0.94 \\
5-10 & 0.062 $\pm$ 0.001 &  0.100 $\pm$ 0.005 & 0.96 \\
10-20 & 0.095 $\pm$ 0.0005 &  0.144 $\pm$ 0.003 & 0.98 \\
20-40 & 0.146 $\pm$ 0.0004  &  0.208 $\pm$ 0.003 & 0.91 \\
40-60 & 0.171 $\pm$ 0.001 & 0.236 $\pm$ 0.006 & 0.86 \\
60-90 & 0.066 $\pm$ 0.001 &  0.091 $\pm$ 0.004 & 1.06 \\
\end{tabular} \end{ruledtabular}
\end{table}
%

\begin{figure}[thb]
\includegraphics[width=1.0\linewidth]{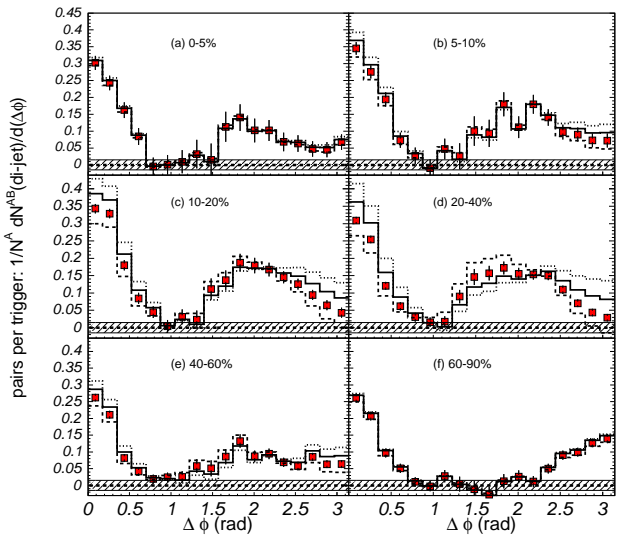}  
\caption{Jet-pair distributions 
$dN^{AB}_{\rm (Di-)Jet}/d(\Delta\phi)$
for different centralities, normalized per
trigger particle. The shaded bands indicate
the systematic error associated with the determination
of $\Delta\phi_{Min}$.
The dashed~(solid) curves are the
distributions that would result from 
increasing~(decreasing) 
$\langle v_{2}^{A} v_{2}^{B} \rangle$ by
one unit of the systematic error; the dotted curve would result
from decreasing by two units.
\label{fig:Subtracted}}
\end{figure}

The average level of the background, $b_{0}$, can,
in principle, be fixed by making an assumption about 
the shape of the (di-)jet pair distribution. 
However, since we wish to measure the shape of the (di-)jet
azimuthal correlations, we use a technique that
requires no such {\it a priori} assumptions.
The simplest assumption allowing $b_{0}$ to be 
fixed is that $dN^{AB}_{\rm (Di-)Jet}/d(\Delta\phi)$ is zero for
at least one value of $\Delta\phi$ ({\em i.e.} $\Delta\phi_{Min}$).
We refer to this as the ZYAM (``zero yield at minimum'') 
assumption for the (di-)jet pair distribution.
The ZYAM condition is met by varying $b_{0}$
until the background component matches a functional fit
to the correlation function at one point $\Delta\phi_{Min}$,
as illustrated by the solid bands in Fig.~\ref{fig:Unsubtracted}.
The systematic error on $b_{0}$ associated with this
procedure (see Fig.~\ref{fig:Subtracted}) was estimated
by using a variety of functional forms that matched
the data.

A non-zero yield of (di-)jet pairs at $\Delta\phi_{Min}$
would invalidate the ZYAM assumption and result in an 
overestimate of the value of $b_{0}$.
To verify that we are not making significant error in the 
normalization of the background, we have independently estimated 
the $b_{0}$ values using the $AB$ pair combinatorial rate,
corrected for a slight bias introduced by mixing events of
different multiplicity within the same centrality 
class~\cite{ppg033}.  These
independent estimates are consistent with or even slightly higher
than the ZYAM-determined values for $b_{0}$, confirming that we
are not significantly over-estimating the background levels.

Once $\langle v_{2}^{A} v_{2}^{B} \rangle$ and $b_{0}$ are fixed, 
we can extract $J(\Delta\phi)$ and the fully corrected 
\mbox{(di-)jet} pairs distribution $dN^{AB}_{\rm (Di-)Jet}/d(\Delta\phi)$.
We construct the conditional yield distribution of 
jet-associated partners per trigger:
\begin{equation}
\frac{1}{N^{A}}
\frac{dN^{AB}_{\rm (Di-)Jet}}{d(\Delta\phi)} =
\frac{J(\Delta\phi)}{\int C(\Delta\phi') \; d(\Delta\phi')}
\; \frac{N^{AB}}{N^{A}}
\label{Eq:Pairs_extract}
\end{equation}
\noindent
Here, $N^{A}$ is the number of triggers and $N^{AB}$ 
the total number of $AB$ pairs in the event sample. 
Assuming that the pair efficiency is the product of 
the single particle efficiencies the trigger ($A$)
efficiency cancels in Eq.~\ref{Eq:Pairs_extract}. 
Thus, the ratio is corrected for acceptance and reconstruction 
efficiency~\cite{ppg023} of the lower-$p_{T}$ $B$
particles; the systematic error on this correction 
leads to a 10\% uncertainty on the associated yields.


The conditional yields of (di-)jet-induced partners per
trigger are shown in Fig.~\ref{fig:Subtracted}.
For the most peripheral event sample the (di-)jet
associated yield distribution has an appearance we
might expect from a normal (di-)jet fragmentation 
process~\cite{ppg029,ppg039}: a well-defined near-side peak around
$\Delta\phi=0$ and a somewhat wider away-side
peak around $\Delta\phi=\pi$.  For more  central event samples 
the shape of the near-side peak is essentially unchanged while 
the associated yield in the near-side peak increases, indicating 
some change in the fragmentation process .  

%
%

The much more dramatic change, however, is in the 
away-side peak which is considerably broader in all the 
event samples more central than 60\%.  In mid-central and 
central collisions there is a local minimum at 
\mbox{$\Delta\phi=\pi$}. The existence of these local minima 
{\it per se} is not significant once we take the systematic
errors on $\langle v_{2}^{A} v_{2}^{B} \rangle$ into
account (see below), but it is clear that the away-side
peaks in all the more central samples have a very 
different shape than in the most peripheral sample.

%
%

Given the dramatic results for the away-side peaks seen 
in Fig.~\ref{fig:Subtracted}, it is important to establish 
that they are not simply artifacts created by our method 
for background pair subtraction.  If we relax the ZYAM 
assumption and lower $b_{0}$ slightly, 
the effect on any (di-)jet pair distribution would 
essentially 
be to raise it by a constant, which would not change the
presence of the local minima at $\Delta\phi=\pi$.

Changes to our estimate for 
$\langle v_{2}^{A} v_{2}^{B} \rangle$ can alter the 
shape of the (di-)jet distribution for some centrality 
samples, but the result of away-side broadening with
centrality remains robust. The curves in 
Fig.~\ref{fig:Subtracted}
show the distributions that would result if the
$\langle v_{2}^{A} v_{2}^{B} \rangle$ products were 
arbitrarily lowered by one and two units of their
systematic error.  With a two-unit
shift the shape in the mid-central would no longer
show significant local minima at $\Delta\phi=\pi$.
However, the widths of the away-side
peaks are clearly still much greater than in the
peripheral sample and the distributions in the two most
central samples are hardly changed at all in shape.
Even lower values of $\langle v_{2}^{A} v_{2}^{B} \rangle$ 
could be contemplated, but they would still
not change the qualitative result of away-side broadening.
And, such low $\langle v_{2}^{A} v_{2}^{B} \rangle$ values
would also require a severe breakdown of the assumption 
$\langle v_{2}^{A} v_{2}^{B} \rangle =
\langle v_{2}^{A} \rangle \langle v_{2}^{B} \rangle$,
indicating that these background pairs have a large,
hitherto-unknown source of azimuthal anti-correlation.

Convoluting the jet fragments' angles with respect 
to their parent partons and the acoplanarity between 
the two partons~\cite{ppg039} would
yield a Gaussian-like shape in $\Delta \phi$, 
possibly broadened
through jet quenching\cite{Qiu_04,Wiedemann}.  
The observed shapes in the away-side peaks cannot 
result from such a convolution.

\begin{figure}[thb]
\includegraphics[width=1.0\linewidth]{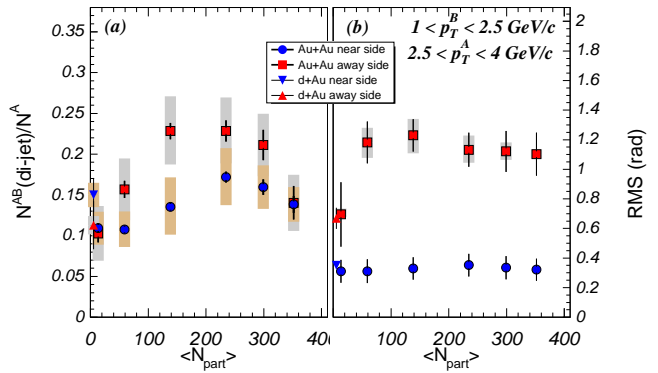}  
\caption{(a)~Associated yields for near- and 
away-side peaks in the jet pair distribution,
and (b)~widths (RMS) of the peaks in the full
0--$2\pi$ distributions; plotted versus
the mean number of participating nucleons for each
event sample.  Triangles show results from 0--20\% 
central d+Au collisions at the same 
$\sqrt{s_{_{NN}}}$ \cite{ppg039}.  
Bars show statistical errors, shaded bands systematic.
\label{fig:Yields_Widths}}
\end{figure}

We define the part of the $\Delta \phi$ distribution in 
\mbox{$|\Delta\phi|<\Delta\phi_{\rm Min}$} as the
``near-side'' peak and  \mbox{$|\Delta\phi|>\Delta\phi_{\rm Min}$}
as the ``away-side'' peak.
Each peak is characterized by 
its yield of associated partners per trigger, and by its RMS width.  
We measure these for the full peak in the distribution over 
all values of $\Delta\phi$; the folded distributions over 
$0<\Delta\phi<\pi$ shown here contain only half of each full peak's shape.
These yields and widths are plotted in Fig.~\ref{fig:Yields_Widths}
for the different Au+Au centrality samples, along with the
same quantities for 0--20\% central d+Au collisions at 
$\sqrt{s_{_{NN}}}$=200~GeV \cite{ppg039}. 
The yields and widths for the near- and away-side peaks
in peripheral Au+Au collisions are consistent with
those in d+Au collisions.  The yields of both the near- and 
away-side peaks increase from peripheral to
mid-central collisions, and then decrease for the most central
collisions.  The near-side width is unchanged with centrality, 
while the away-side width increases substantially from the 60--90\% 
sample to the 40--60\% sample and then remains constant with 
centrality.
%
%
%
%
%
%

In summary, we have presented correlations of high momentum
charged hadron pairs as a function of collision centrality in
Au+Au collisions.  Utilizing a novel technique we extract
the jet-induced hadron pair distributions and show that the 
dense medium formed in Au+Au collisions at RHIC modifies jet 
fragmentation.  
In central and mid-central collisions the away-side angular 
distribution is significantly broadened relative to peripheral 
and d+Au collisions, and appears to be non-Gaussian.  
The shapes of the away-side $\Delta\phi$ distributions 
for non-peripheral collisions are apparently not consistent 
with purely stochastic broadening of the peripheral Au+Au away-side. 
However, the broadening and possible 
changes in shape of the away-side jet are suggestive of 
recent theoretical predictions 
of dense medium effects on fragment distributions 
\cite{cherenkov_1,cherenkov_2,cherenkov_3,armesto_04b}.
The broadened shapes of the away-side distributions also imply 
that integration of the away-side peak in a narrow angular range around 
$\Delta\phi=\pi$ yields fewer associated partners 
in central collisions
than in peripheral/d+Au collisions, as seen elsewhere\cite{starb2b,ppg033};
but integrating over the entire broadened peak recovers the 
jet partners in the range 1.0~\gevc~$<p_{T}^{B}<$~2.5~\gevc $ $
used here.  Even though  two-particle correlations do 
not allow for full reconstruction of the jet fragmentation function, 
these data provide an entirely new way to probe the hot, 
dense medium formed in heavy ion collisions.

\vspace{1.0cm}


\begin{acknowledgments}
We thank the staff of the Collider-Accelerator and Physics
Departments at BNL for their vital contributions.  We acknowledge
support from the Department of Energy and NSF (U.S.A.), 
MEXT and JSPS (Japan), CNPq and FAPESP (Brazil), NSFC (China), 
CNRS-IN2P3 and CEA (France), 
BMBF, DAAD, and AvH (Germany), 
OTKA (Hungary), DAE and DST (India), ISF (Israel), 
KRF and CHEP (Korea), RMIST, RAS, and RMAE (Russia), 
VR and KAW (Sweden), U.S. CRDF for the FSU, 
US-Hungarian NSF-OTKA-MTA, and US-Israel BSF.
\end{acknowledgments}



\end{document}